\documentclass{article}
\usepackage[a4paper, total={6in, 8in}]{geometry}

\usepackage{graphicx}
\usepackage[utf8]{inputenc}
\title{Visual Exploration and Energy-aware Path Planning via Reinforcement Learning}

\date{\vspace{-10ex}}
\begin{document}
\maketitle
\begin{center}

\author{Amir Niaraki, Jeremy Roghair, Ali Jannesari }

\vspace{\baselineskip}
$\{$niaraki, jroghair, jannesari$\}$@iastate.edu
\vspace{\baselineskip}

Department of Computer Science, Iowa State University

% \date{January 2021}
\end{center}

\begin{abstract}
Visual exploration and smart data collection via autonomous vehicles is an attractive topic in various disciplines. Disturbances like wind significantly influence both the power consumption of the flying robots and the performance of the camera. We propose a reinforcement learning approach which combines the effects of the power consumption and the object detection modules to develop a policy for object detection in large areas with limited battery life. The learning model enables dynamic learning of the negative rewards of each action based on the drag forces that is resulted by the motion of the flying robot with respect to the wind field. The algorithm is implemented in a near-real world simulation environment both for the planar motion and flight in different altitudes. The trained agent often performed a trade-off between detecting the objects with high accuracy and increasing the area coverage within its battery life. The developed exploration policy outperformed the complete coverage algorithm by minimizing the traveled path while finding the target objects. The performance of the algorithms under various wind fields was evaluated in planar and 3D motion. During an exploration task with sparsely distributed goals and within a UAV's battery life, the proposed architecture could detect more than twice the amount of goal objects compared to the coverage path planning algorithm in moderate wind field. In high wind intensities, the energy-aware algorithm could detect 4 times the amount of goal objects when compared to its complete coverage counterpart. 

\vspace{\baselineskip}

\noindent \textbf{keywords:} Path planning, Reinforcement learning, Object detection, Unmanned aerial vehicles, Energy-efficiency.
\end{abstract}

\section{Introduction}

Unmanned Aerial Vehicles (UAVs) are employed in variety of disciplines, for applications from vegetation detection in large farms to target search and rescue \cite{sun2016camera}. Today, although UAVs are popularly in use, they suffer from limited battery life. Sufficient aerial imagery in large fields is typically achieved by multiple drone flights, which are often performed by complete coverage of the domain. Wind plays the most significant role in power consumption of aerial vehicles. It is shown that, by only changing the yaw of a quadrotor with respect to wind vector, we can improve the covered path by 30$\%$ on the same battery life \cite{vasisht2017farmbeats}.  Therefore, path planning of the UAVs and monitoring of power consumption with respect to wind is an attractive research topic during  autonomous task completion missions.

\begin{figure}[t]
  %\scalebox{.5}{\input{plot.tex}}
  \centering
  \includegraphics[width=0.8\textwidth]{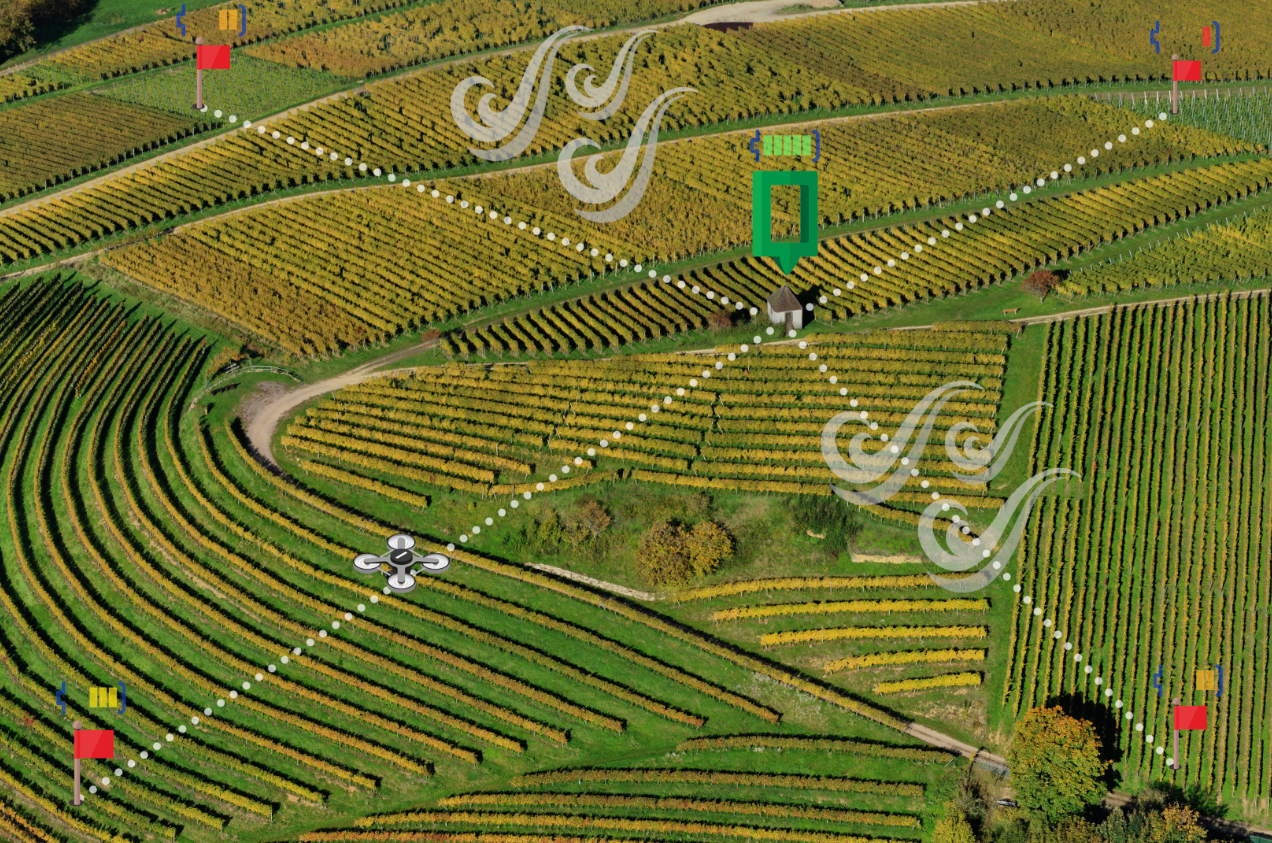}
  \caption{Adapting to wind condition can significantly reduce the power cost of UAV system in large fields with sparse goals.}
  \label{fig: DroneInFarm}
%  \FloatBarrier
\end{figure}

There are two folds to UAV motion control while addressing the completion of a Task. Firstly, flight control inherently implies stabilization and position control of an aircraft, which is executed by an onboard Flight Control Unit (FCU) in an “inner loop” level. Secondly, a control unit in “outer loop” level, is typically responsible for mission level objectives such as path planning, collision avoidance and navigation~\cite{koch2019reinforcement}. Let us define the problem of periodic (i.e. daily) crop monitoring by covering the farm demonstrated in Fig.~\ref{fig: DroneInFarm}. The task is to provide aerial imagery in the sparse locations in the field, frequent enough to reliably monitor the health of the crops, but not through redundant visits, which results in power depletion of the drone. While wind can assist the drone to reach certain far spots, it may increase the power cost of taking other paths. Importantly, disturbances like wind are time varying and often there is no model for the behavior of such natural objectives in-hand. Therefore, tackling such problems requires adaptive path planning and careful design for goal prioritization algorithms. In problems such as what is shown in Fig.~\ref{fig: DroneInFarm}, one approach is to evaluate the planned path at certain intervals via an implemented wind-power model in the control unit.Thus far, Bezzo et al. have comprehensively studied such goal scheduling problem under wind, with a model predictive  control algorithm~\cite{bezzo2016online} in simulation and lab experiment. A benchmark study on energy-aware coverage path planning of UAVs by Di Franco and Buttazzo\cite{di2015energy} have tackled the path planning problem to minimize energy consumption while satisfying mission requirements. However, model-based estimation of power cost cannot be reliably generalized across different robots.
Considerations on the objective is vital for optimal operation of autonomous UAV agent. Among other components, the performance of the computer vision module normally plays a substantial role in vision-based exploratory missions~\cite{vaddi2019objectDetection}. Therefore, it is preferred to utilize a framework, which simultaneously addresses the objective requirements and the power constraints. Thus, we employed a reinforcement learning (RL) approach to utilize the UAV's  interaction with its environment for collecting sufficient power consumption information concerning varying wind fields while rewarding the robot by the detected objects so that it strives to explore.

Notably, Li et al. have proposed a Q-learning approach for the task of search for an optimal path in goal approaching~\cite{li2006q}. They followed the course coding paradigm in RL, where the search domain is broken down to a pixelated grid-world. More recently, the case of grid path planning with presence of obstacles was addressed with deep reinforcement learning \cite{panov2018grid}. Our work follows the classic course-coding paradigm and reduces the search domain to a grid-world along the work of Le et al., but with the goal of tackling the power constraint problem for UAV path planning. The effect of power constraints in the control of multiple UAVs for the exploration of the environment was studied by Liu et al. \cite{liu2018energy} using available power models.Quadrotors suffer from their short flight duration capacity, and wind is the determinant factor in their power consumption. Therefore, we are proposing a Q-learning based path planning and goal scheduling framework which does not operate on a preexisting power model but interacts with its environment to dynamically determine its action cost at every step. This framework follows the approach that is regarded as learning from model or End-to-End training category of robot reinforcement learning\cite{wang2020deep}, which is shown to bear a superior performance compared to the frameworks which operate via separate hand-engineered components for perception\cite{mammadli_ea:taco:2019}, state estimation\cite{asli2020energyaware}, and low-level control\cite{levine2016end}. In the benchmark study by Levine et al.,\cite{levine2016end} it is demonstrated that, training the perception and control systems jointly end-to-end provide better performance than training each component separately.       

 In the studied case here, the flying agent starts with no knowledge of its power cost function, disturbance information, and behavior of objective function and achieves a path planning policy through solving a power optimization problem through interacting with a simulation environment. RL algorithms based on state-action value tables have been utilized to address the path planning  problem. To the knowledge of the authors, this is the first time that RL is used for goal-selection and path planning on varying wind conditions. The developed model is evaluated in various scenarios where the agent is required to find randomly distributed target goals in a search domain, and its performance in planar and 3D motion is demonstrated. Particularly, for the first time we tackle the problems with large search domains, where the model is required to generate a path planning policy to find sparsely distributed goals and prioritize its targets in a disturbance-heavy environment.  
 
The rest of this article is organized as follows: Section II presents a high-level description of UAV flight dynamic and general RL framework. In Section III, the proposed framework is described besides further details on the design of RL-based path planning framework, such as effective random search functions, how to tune the trade-off between the punishment of each step and the reward of detection. The simulation results and a discussion on its expansion to real-world cases is given in Section IV and the closing remarks are provided in Section V.

\section{Overview}
Here, a generalized explanation for quadrotor flight dynamic is provided to create the background required to understand this work. Next, a brief overview of reinforcement learning (RL) and two commonly known RL algorithms: Q-learning and SARSA are presented. 

\subsection{UAV  flight dynamic}

The UAV flight dynamic in this study is simplified to a planar motion with 3 degrees of freedom which includes position in x and y axis and the heading angle $\psi$ while the altitude of the quadrotor (position in z axis) remains constant. It is reasonable to consider the quadrotor as a rigid body, which accelerates by the torques and forces applied form its four rotors. The velocity and applied wind is simplified in order to reduce the complexities derived by the physics of the UAV, and shown schematically in Fig.~\ref{fig: droneschematics}. For the given velocities, we can have:
\begin{figure}[thpb]
  %\scalebox{.5}{\input{plot.tex}}
  \centering
  \includegraphics[scale=0.4]{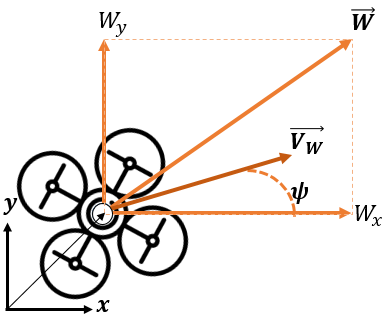}
  \caption{Kinematic relation of quadrotor velocity vs wind vector.}
  \label{fig: droneschematics}
%  \FloatBarrier
\end{figure}

\begin{equation} %\tag{2.0}
     \dot{X_G}=V_w cos(\psi)+W_x
%    \label{eq: 1}
\end{equation}
\begin{equation} %\tag{2.0}
     \dot{Y_G}=V_w sin(\psi)+W_y
%    \label{eq: 1}
\end{equation}
\begin{equation} %\tag{2.0}
     \dot{\psi}=\frac{V_w}{R_{min}}U\quad  (-1<U<1)
%    \label{eq: 1}
\end{equation}

These equations can be integrated with respect to time to give:
\begin{equation} %\tag{2.0}
     X_G=-\frac{R_{min}}{U}cos(\psi_{0}+\frac{V_w}{R_{min}}Ut)+{W_x}{t}+X_{G_{0}}
%    \label{eq: 1}
\end{equation}

\begin{equation} %\tag{2.0}
     Y_G=-\frac{R_{min}}{U}sin(\psi_{0}+\frac{V_w}{R_{min}}Ut)+{W_y}{t}+Y_{G_{0}}
%    \label{eq: 1}
\end{equation}

\begin{equation} %\tag{2.0}
     \psi=\psi_0+\frac{V_w}{R_{min}} UT
%    \label{eq: 1}
\end{equation}

Where, the variables $\dot{X_G}$ and $\dot{Y_G}$, represent the UAV’s total velocity in the x and y direction respectively, relative to the ground. $W_x$ and $W_y$ are the wind speeds in the x and y directions, respectively. $\psi$  gives the angular velocity $R_min$  represents the UAV’s minimum turning radius and $V_m$ represents the relative velocity of the UAV. Finally,  $X_G$, $Y_G$ gives global $x$ and $y$ coordinate of the UAV while $\psi$ describes its heading angle\cite{al2012wind}. 

\subsubsection{Drag Coefficient}
In order to find the power consumption of the quadrotor, the fluid dynamic of the wind flow around the rigid body is modeled to give the drag coefficient, calculated numerically by Ansys Fluent~\cite{yue2013ansys}. Drag coefficient is defined by Eq. \ref{eq: dragcoefficient}. Where $F_d$ is the drag force, $\rho$ is the mass density of air, $V_w$ is the velocity of the drone relative to the fluid and $A$ is the surface area:
\begin{equation}
    c_d=\frac{2F_d}{\rho {V_w}^2 A}
    \label{eq: dragcoefficient}
\end{equation}

\begin{figure}[h]
  %\scalebox{.5}{\input{plot.tex}}
  \centering
  \includegraphics[width=0.7\textwidth]{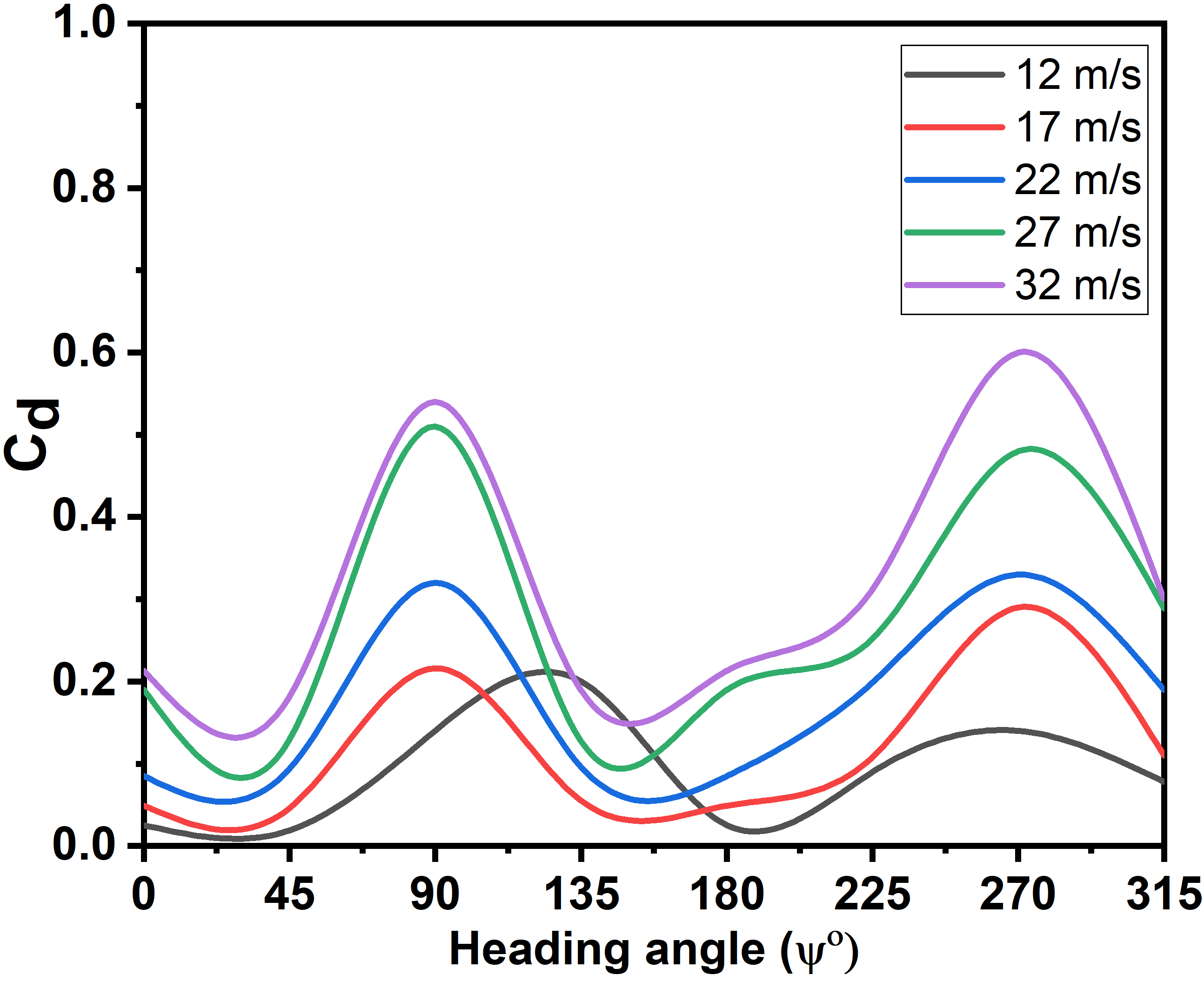}  
  \caption{The change in heading angle ($\psi$) relative to the magnitude and direction of the wind vector results in highly varying drag forces in the body of the vehicle.}
  \label{fig: DragTheta}
%  \FloatBarrier
\end{figure}

In all the simulations here, the quadrotor velocity was set to $22 m/s$ and the absolute value of the wind speed was set to $|W|\in\{-10,-5,0,5,10\}\; m/s $. The details on mathematical modelling and parameter identification of quadrotor can be found in the Ref.~\cite{chovancova2014mathematical}. Briefly, the drag coefficient represents a dimension-less measure for the drag force that is applied to a particle when moving in a fluid. In our case, the direction and speed of the wind with respect to the quadrotor's movement determines the amount of power consumed from its battery at each moment.  

the drag coefficient of the drone in various wind vectors was calculated in 8 different $\theta$s which will be further used for calculating the power cost of each taken action by the agent in the simulation environment. The data from Figure \ref{fig: DragTheta} was used as a database for estimating the power consumption of the agent at each step. Each line represents the relative velocity of the drone with respect to the headwind. 

%------------------------------------------------------------------------

\subsection{Reinforcement learning: Q-learning and SARSA}
\begin{figure}[thpb]
  %\scalebox{.5}{\input{plot.tex}}
  \centering
  \includegraphics[width=0.7\textwidth]{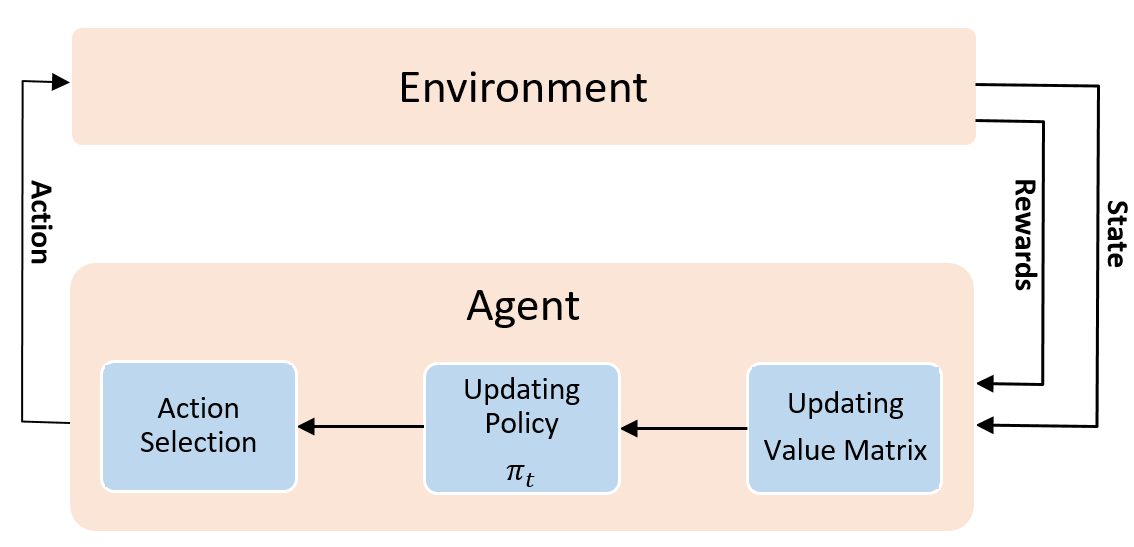}
  \caption{Standard network structure for reinforcement learning algorithm.}
  \label{fig: RLschem}
%  \FloatBarrier
\end{figure}

Robot reinforcement learning is an increasingly popular method that offers the capability of learning the previously missing abilities. These can include behaviors that are priory unknown, are not facile to code, or optimizing problems without an accepted closed solution \cite{zamora2016extending}.
The behavior optimization occurs through repetitive trial and error interaction between an agent and its environment. This machine learning method can be defined as a Markov decision process (MDP) through which the agent is trained by an action-sense-learn cycle \cite{chen2012evaluating}. In a standard model-based RL algorithm (Fig.~ \ref{fig: RLschem}), the agent observes the state $s_t\in S$ from its environment, and takes action $a_t\in A$ based on the prior knowledge resulting in its current policy $\pi_t$. The taken action results in a new state $s_{t+1}$, which here can be determined from the state transition distribution $P(s_{t+1} |s,a)$ and leads to the reward $r(s,a)$. 

In the solved cases here, generally the UAV agent receives update on its location and velocity at each state (x-y-z location). Based on the current policy $\pi_t$ it takes an action by noting the value of the experienced state-action pairs $(s,a)$. Once reached to a new state, a reward value will be calculated by the model based on the agent movement cost and potential accomplished goals.  The expected return (sum of discounted rewards) can consequently be used to give the optimal state-action value function for a given state-action pair $(s,a)$:

\begin{equation}
Q^*(s_t,a_t) = r(s_t,a_t)
+\gamma\Sigma _{s_{t+1}\in S} P(s_{t+1}|s_t,a_t)\max _{a_{t+1}\in A} Q^*(s_{t+1},a_{t+1})
\end{equation}

Where t can be an iteration numerator (or time-step), $\gamma \in (0,1)$ is a pre-defined discount factor. Therefore, the agent learns to modify his action policy based on the cumulative rewards over iterations. The agent’s policy is essentially a mapping from each state to its corresponding action. Using this state-action value function, we can calculate the optimal policy, $\pi^*$ by:
\begin{equation}
    \Pi^* (s,a)=argmax_{a_t\in A}Q^*(s_t,a_t)
\end{equation}

Various RL algorithms mostly vary in terms of trade-off between exploration and exploitation in creating and updating the value function \cite{sutton2018reinforcement}. Here we will describe Q-learning and SARSA and further implement them in the experimental scenarios. 
Q-learning is an Off-Policy algorithm for temporal difference (TD) learning. While not requiring a model of the environment, based on an exploratory or random policy, Q-learning learns to optimize the policy when the actions are selected. In Q-learning, the learned action value function, Q, directly approximates $Q^*$ independent of the followed policy:

\begin{equation}
    Q(s_t,a_t)\leftarrow Q(s_t,a_t)+\alpha [r_{t+1}+
    \gamma max_a Q(s_{t+1},a_{t+1})-Q(s_t,a_t)]
\label{eq: QQQ}
\end{equation}

Where, $\alpha \in (0,1)$ is the learning rate, which is a hyper-parameter to tune the significance of the most recent rewards.

SARSA is an On-Policy temporal difference (TD) learning method which uses the following equation to update its action value function, Q:

\begin{equation}
    Q(s_t,a_t)\leftarrow Q(s_t,a_t)+\alpha [r_{t+1}+
    \gamma Q(s_{t+1},a_{t+1})-Q(s_t,a_t)]
\label{eq: sars}
\end{equation}

The major difference between SARSA and Q-learning lies upon the choice of action in each state. SARSA uses every element of these five events: $(s_t, a_t, r_{t+1}, s_{t+1}, a_{t+1}$) that creates the transition from one state-action pair to the next based on the current policy. In contrast the choice of action for the Q-learning algorithm, is normally performed by an $\epsilon$-greedy approach. Where for a small probability of $\epsilon$ a random action is chosen to ensure exploration in the state-space \cite{sutton2018reinforcement}. For the actions, where the probability of $\epsilon$ is not triggered, the agent will take action that would maximize the reward based on the updated Q-matrix.

\subsection{Simulation environment}
\begin{figure}[h]
  %\scalebox{.5}{\input{plot.tex}}
  \centering
  \includegraphics[width=15cm]{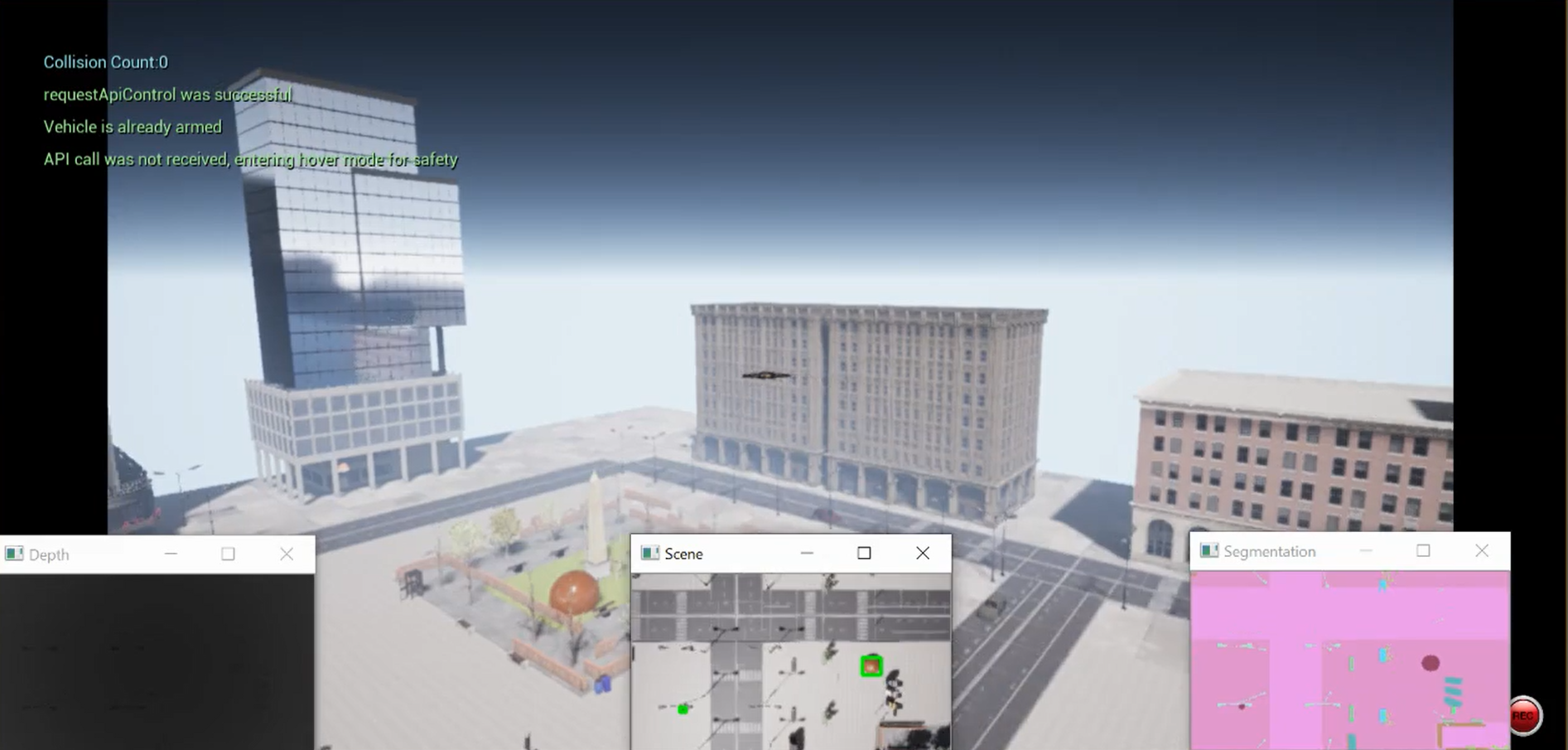}  
  \caption{Simulation environment in Unreal Engine with the detections depicted on image stream. Please refer to this link for the demo: https://youtu.be/kea1sEz9NVE}
  \label{fig: Simulation}
%  \Pl
\end{figure}

A major challenge for Rl-traning in robot motion planning is the need for numerous trials, many of which, resulting in termination of the learning episode by an unwanted actions leading to  failure of the mission. Therefore experimentation in real world is often initialized by the policy which is already generated in a simulation through what is often regarded as mental rehearsal for the robot \cite{zamora2016extending}.  
In order to enable training of an autonomous UAV in a near real-world environment, the proposed framework is implemented in Unreal Engine~\cite{sanders2016introduction} using Microsoft AirSim \cite{shah2018airsim} API. Unreal Engine is a gaming engine that provides a platform for replicating realistic scenarios and leveraging the advantage of high definition visual details for the computer vision modules of the learning algorithm. Airsim acts as a medium to enable the interaction between the control algorithm and the simulation environment (Fig.~\ref{fig: Simulation}).

%-----------------------------------------------------------
%-----------------------------------------------------------
\section{Methodology}

\begin{figure*}
\begin{center}
  \includegraphics[width=0.8\textwidth]{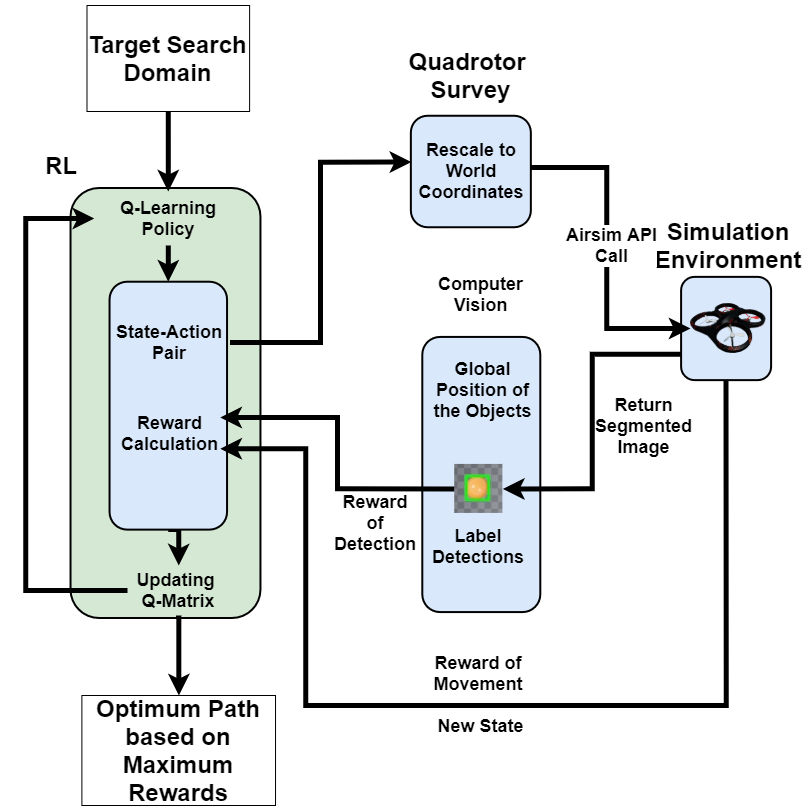}  
\end{center}
   \caption{Workflow for autonomous path planning using object detection vision and reinforcement learning.}
\label{fig: workflow}
\end{figure*}

Operation of autonomous agents in large areas often bears an inherent goal exploration problem. RL is a desirable paradigm especially where there is no explicit model for the distribution of goal in hand. Due to the necessity of numerous trials for obtaining a behavioral policy, robustness and broad applicability of a RL control unit is essential. The presented framework here, operates on the classic state-action value matrix. However, characteristics of the agent and the target are desired to be finely reflected in the updating process of the Q-matrix. Fig. \ref{fig: workflow} demonstrates the architecture for the communication of RL and computer vision module with the simulation environment. 
\subsection{Reinforcement learning module}
In order to define the problem in the RL framework, the entire domain was broken down to a grid world with the course coding technique\cite{sutton2018reinforcement}. The state of the agent is defined by its position on the grid,  $s_x\in[0,WorldWidth]$, $s_y\in[0,WorldHeight]$, $s_z\in[0,WorldAltitude]$,and its battery level $s_b\in(0,100)$ while the environment imposes $w_x\in\{-W_{max}, -W_{max}/2, 0, W_{max}/2, W_{max}\}$. The resolution of this discretization process can directly affect the accuracy of the whole framework and has been chosen as one of the main parameters of study.

%--------------------------------------------------------------------------
\begin{figure}[h]
  %\scalebox{.5}{\input{plot.tex}}
  \centering
  \includegraphics[width=0.5\textwidth]{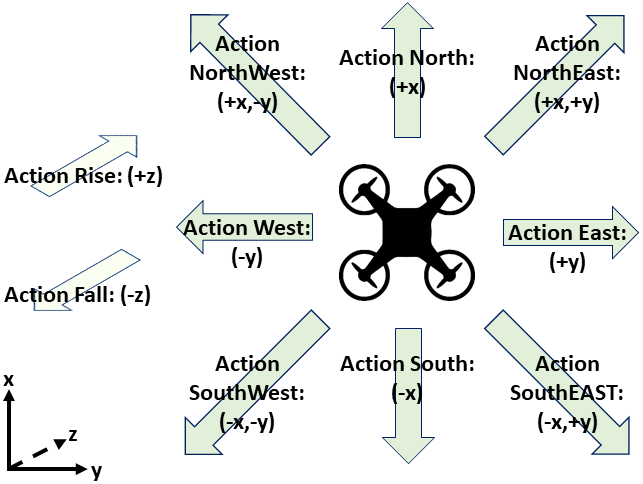}  
  \caption{The list of 10 actions that the agent can take to move laterally or change its altitude.}
  \label{fig: DroneActions}
%  \FloatBarrier
\end{figure}

%--------------------------------------------------------------------------
At each state, the agent can choose from a list of 10 actions given in Fig. \ref{fig: DroneActions}, where the first 8 transfer its state to the adjacent 8 lateral states, and action  9 and 10 increase or decrease its altitude. The actions which both enable change in altitude during lateral motion are not added to the action list to prevent the creation of an extraordinary large state-action space. 

Each episode is initialized at a predetermined start state and the battery level is set to its maximum ($s_b=100$). After choosing an action $a_t$ from the valid action list the power cost of traveling between the two states ($s_t, s_{t+1}$) will be calculated based on the drag coefficient, drone speed and traveled distance between the two states to assign a step reward between each two steps, presented as $r_{movement}$. As the drone traverses between the two steps, the computer vision module returns the number of the newly detected objects and assigns a reward of detection, $r_{detection}$ to ($s_t$,$a_t$). The two rewards will be summed with the tuning parameter, $C_r$ given by Eq. \ref{eq: TuningParameter} to constitute the overall reward $r_t$ of step $s_t$. This value will be used to update the Q-matrix given in Eq. \ref{eq: QQQ}.  
\begin{equation}
    r_{movement}=-PowerCost(s_{t+1}-s_t)
\end{equation}
\begin{equation}
    r_t=r_{movement}+r_{detection}\times c_r
    \label{eq: TuningParameter}
\end{equation}
\begin{equation}
    s_{b_{t+1}}=s_{b_{t}}-PowerCost(s_{t+1}-s_t)
\end{equation}

The role of $C_r$ is to tune the trade-off between the object detection goal and the traveling cost. A very high value of $C_r$ can result in a spike in the Q-matrix which may demotivate the agent to explore for other goals, while a very low value of $C_r$ may lead to redundant steps which are solely concerned about minimizing the power cost. For instance, the $r_{movement}$ will acquire a value of -18.5 at its lowest (when moving with a headwind at the simulations maximum speed), $C_r$ is set to $50$ and the $r_{detection}$ is an integer equivalent to number of detected balls. In order to prevent the agent from leaving the domain, the Q-matrix is initialized with a set of conditions which assign the value of undesired actions to -100 in the edge states. An additional negative reward for recharging was set to -30 upon the battery charge $(s_b=s_{b_{max}})$ for the scenarios with a charging spot. 

It is assumed that the UAV moves in  constant speed. Thus, $s_b$ is dependent on the change of location in consecutive time-steps and wind vector at that location and time-step. Consequently, the agent can receive updates on its battery level solely by referring to the power use during each step based on its relative speed with respect to wind corresponding to the drag coefficient.

One common problem in the path planning with Q-learning is the emergence of repetitive steps between two or more consecutive states. For instance, if the agent find an object at state $s$ it will in high value of the actions leading to state $s$, and thus these values propagate to the adjacent states. Then the choice of highest rewarding state will constitute an oscillatory movement between the neighbors. In order to prevent such patterns, for each state a list of valid actions are created in the RL module and once the agent experienced the state-action pair $(s_t,a_t)$ this pair will be removed from the valid choices in the Q-matrix for all time steps after $t$ until the end of the episode. This may lead to situations where there may be no valid actions left for a state in corners and edges, which results in a reward of -200 and the termination of the episode.

\begin{equation}
\epsilon_{episodic}=(\epsilon_{init})^\frac{e/E}{1-e/E}
\label{eq: episodic}
\end{equation}

In order to initialize Q-learning, value update was performed through taking random actions with the propability of $\epsilon$ for each wind scenario to train the RL planner, while $\epsilon$ is a constant value. After the first convergence, the RL planner was trained on all wind fields until convergence using an exponentially decaying $\epsilon_{episodic}$ given in Eq. \ref{eq: episodic}.
\subsection{Computer vision module}

Three image streams from the quadrotor's bottom-center camera, i.e., Scene, Segmentation, and Depth Perspective were retrieved via Airsim. The segmentation view delivers a frame consisting of the objects assigned with a specific color in the range 0 - 255. A color segmentation methodology was employed on the segmented image for locating positive detections.

The simulation in Unreal Engine consists of a city environment, quadrotor, and orange balls of variable number and sizes distributed across the domain as the goal objects. The landscape is divided into 10x10 components, and each of these components is sub-divided into quads. The size of the quad is mutable, and the value is chosen based on the experimentation requirements.These objects are assigned a unique ground truth segmentation ID upon detection once the simulation commences. To avoid camera disorientation during quadrotor's rapid motions the Gimbal was stabilized with a fixed Pitch=270$^{\circ}$, Roll=270$^{\circ}$ and Yaw=0.

The experiment consists of three processes; RL algorithm (parent process), quadrotor survey (child process), and segmentation (child process), which run in synchronization with the help of shared variables in the memory (see Fig.\ref{fig: workflow}) . The RL algorithm runs across a specified number of episodes passing the state pair as a tuple of the current state and the next state, resulting in an action. This state pair is passed as an input to the child processes and rescaled to coordinates corresponding to Unreal Engine for simulating the quadrotor's flight.

The segmentation process is responsible for identifying the goal objects, labeling them to avoid repetitive detections and assign rewards to the current state. A mask is applied on the segmented image to include only the pixels of the object of interest and then compute the centroid and bounding box coordinates from the connected components. The detections are depicted in the Scene view by enclosing them in a bounding box using these coordinates. The labels are the global locations assigned to the goal objects ($O_x, O_y$) calculated from the quadrotor position $(X_G, Y_G, Z_G)$, centroid coordinates of the goal object in the image $(x,y)$, center coordinates of the image $(c_x,c_y)$, and a precalculated focal length $(f_x,f_y)$. It should be noted that the segmented image size is (256,144), which is fixed. The relation between these variables in term of global label for the objects is given as:
\begin{equation}
    O_x = {X_G}\times{x_{scale}} + (x_f - c_x) (-{Z_G} {z_{scale}}) / f_x
\end{equation}
\begin{equation}
    O_y = {Y_G}\times{y_{scale}} + (y_f - c_y) (-{Z_G} {z_{scale}}) / f_y
\end{equation}

%------------------------------------------------------------------------------

%------------------------------------------------------------------------

The environment in each episode generates multiple goals, $g_i$ which are located at $(x_i,y_i)$, where $i\in[1,N_g]$. Termination occurs in either of four conditions: (1) The agent finds all the goals resulting in a reward of 100. (2) The agent runs out of battery $(s_b=0)$ with the reward of -100. (3) The agent takes an action that cause the drone to entirely leave the domain, which is allowed, but will result in the reward of -100. (4) The agent is left with no valid actions. 
%-------------------------------------------------------------------------
%-------------------------------------------------------------------------

\section{Evaluation and Results}

Four scenarios are solved here summarized in Table \ref{tab:1}. In the baseline scenario the RL planer can only operate in planar motion ($x,y$ axes). Nevertheless, minor altitude changes may occur for the drone stabilization purposes. These altitude changes may result in new detection instances which will be accounted for, in the policy development. A small $5\times5$ state space with four goal objects are designed to demonstrate path planning where an optimal path is manually determinable. Scenario 2, evaluates the ability of the agent for finding sparsely distributed goals which has broad applications in search and rescue missions~\cite{maciel2019online} and vegetation/pest detection in large farms~\cite{torres2015automatic}.    

\begin{table}[]
    \centering
        \caption{The Reinforcement learning path planning model was evaluated through 4 scenarios which are solved for 5 various wind intensities.}
        
      \includegraphics[width=15cm]{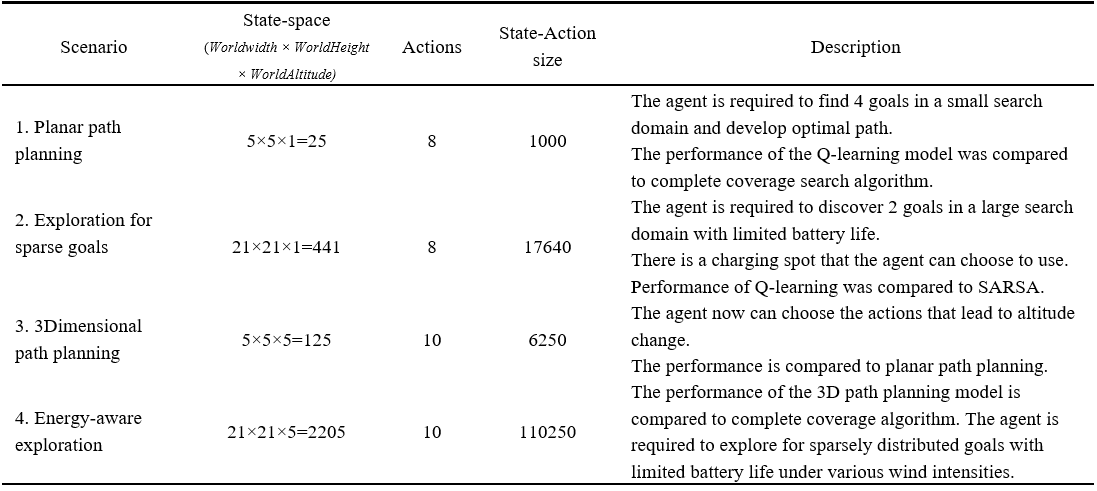}
     \label{tab:1}
\end{table}

In the scenarios 3 and 4, the state-space is expanded vertically and the agent can choose to change its altitude. This can include the effect of the performance of the computer vision module on the path planning. The upper bound of the motion along $z$ axis is set high enough so that the vision module fails to detect some of the smaller objects. In many cases the wind is extreme enough to prevent the possibility for the full coverage of the domain. 
In all of the following experiments the discount factor and learning rate in the RL modules were set to $\gamma=1$ and $\alpha=0.5$ respectively.

%-----------------------------------------------------------------------
\subsection{Planar path planning}
\begin{figure}[h]
  %\scalebox{.5}{\input{plot.tex}}
  \centering
  \includegraphics[width=7.5cm]{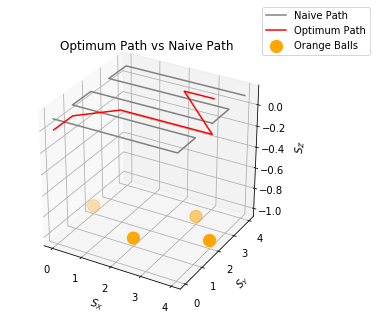}
  \includegraphics[width=7.5cm]{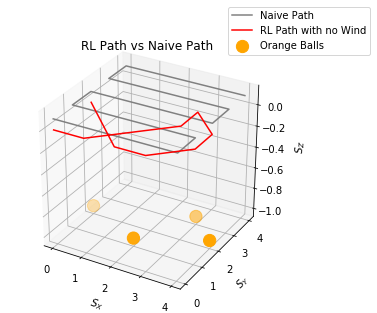}  
  \caption{A comparison between the optimum path generated by the RL planner compared to a complete complete Coverage Algorithm. Wind direction is along the x axis(left), no Wind (right)}
  \label{fig: OptimumvsNaive}
%  \FloatBarrier
\end{figure}

 In this task, the agent is required to initially, explore the environment to find all the objects and then generate the path with minimum required power consumption for capturing the image of all objects. In contrast, the traditional algorithms will perform a sweeping path for complete coverage of the environment (Fig. \ref{fig: OptimumvsNaive}). In order to enable a tangible comparison of the algorithms, the complete coverage path is set to connect the center-point of all states. Nevertheless, providing a thorough aerial image of a domain, typically requires a sweep with image overlap of $40\%-80\%$~\cite{lyu2017autonomous}. The $WorldHeight, WorldWidth$ and $WorldAltitude$ are set to 5,5 and 1 respectively which combined with 5 wind fields result in a state-action space with the size of 1000. The environment is initialized with 4 goals ($N_g=4$) in random spots. All the objects are intentionally placed at the center-point of the states to guarantee their detection by coverage algorithm upon its visit to the object's state object's state. 
 \begin{figure}[h]
  %\scalebox{.5}{\input{plot.tex}}
  \centering
  \includegraphics[width=10cm]{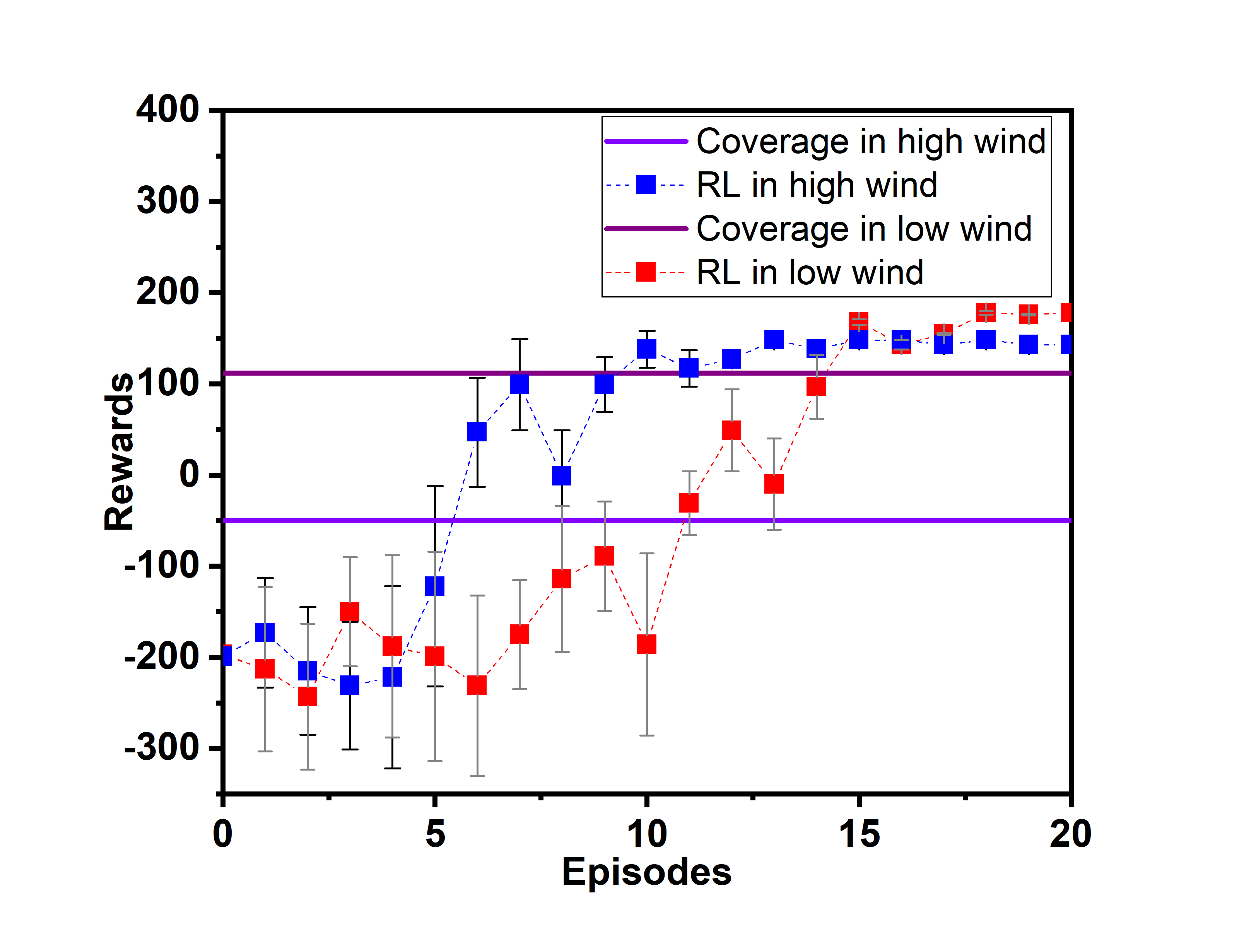}
  \caption{A comparison between the obtained reward per episode by RL planner and complete coverage in large areas with sparse goals. }
  \label{fig: rewards_comparison}
%  \FloatBarrier
\end{figure}

It can be shown that a small state-space with the size of 25, requires 29 steps for a complete coverage, which its corresponding power cost depends on the wind speed and direction. Figure \ref{fig: rewards_comparison} shows that in a high wind scenario the Naive planner runs out of battery just after detecting the first ball resulting a constant reward of -50. In contrast, the RL planner generates the Q-matrix according to its power use during the first episodes. Upon the detection of the goal object via the computer vision module, the reward of detection increases the value of the state-action pairs leading to the goal object and the path convergence toward optimality. For the cases with low wind speed or no wind conditions, the negative rewards of every step often do not force the agent to reduce its number of steps and the agent will not strive to minimize its overall taken steps which can be adjusted by reducing the $C_r$ in Eq.(\ref{eq: TuningParameter}). It is observable that after 10 learning episodes, the RL planner is capable of finding all the balls and maximize its obtained reward. In contrast, the presence of high wind can lead to complete power depletion during the mission, if the complete coverage path is taken. 

\subsection{ Exploration for sparse goals}
\begin{figure}[p]
  %\scalebox{.5}{\input{plot.tex}}
  \centering
  \includegraphics[width=10cm]{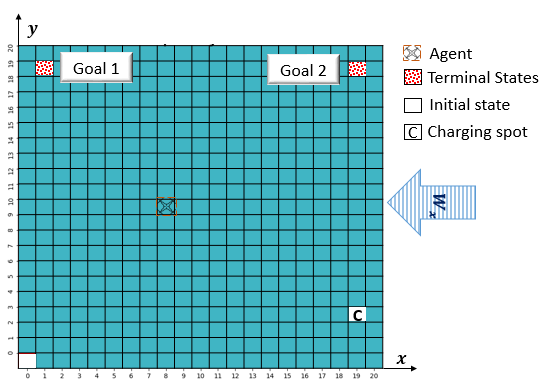}
  \caption{The grid world equivalent for scenario 2: exploration for sparse goals.}
  \label{fig: baseline}
%  \FloatBarrier
\end{figure}

\begin{figure*}[p]
  %\scalebox{.5}{\input{plot.tex}}
\begin{center}
  \includegraphics[width=15cm]{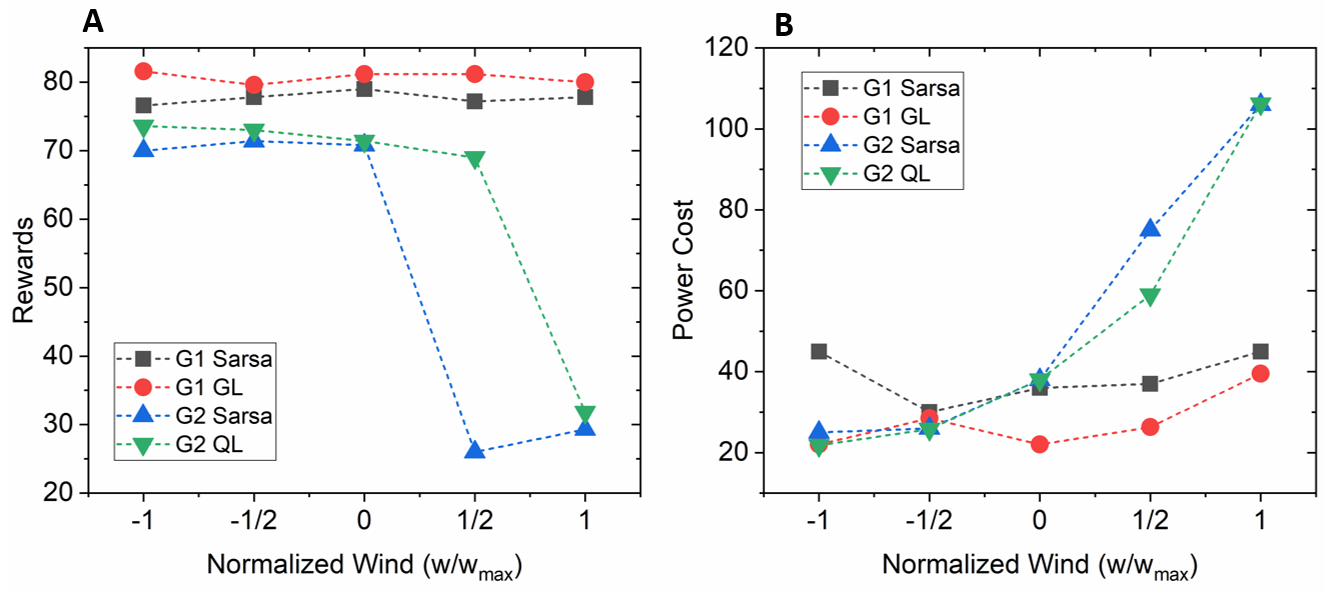}
  \caption{Performance of trained Q-learning and SARSA algorithms for baseline scenario in various constant wind fileds. A) Average rewards after termination by successfully completing the task. B) Average Power cost after termination by successfully completing the task. }
  \label{fig: grid_prior}
%  \FloatBarrier
\end{center}
\end{figure*}

If there are more than one path to be taken, the agent should find the most power-efficient strategy. The initial state is kept constant across the episodes, and the goals $(N_g=2)$ are spread in sparse locations of the domain far from the initial state each are required to be visited only once $c=1$ (Figure.~\ref{fig: baseline}). The agent starts in the initial state with full battery, aiming to first, discover these goals and second, find the path that minimizes its power consumption under the applied constant wind vector $(W(x,y,t)=const)$. In order to, evaluate the ability of the agent to adapt to the conditions; there exists an additional charging spot which might be beneficial to visit depending on the experienced power loss in severe wind. For the ease of representation, 5 constant wind intensities were applied: $w_x\in[-W_{max}, -W_{max}/2,0, W_{max}/2, W_{max}], w_y=0$.  
The ability of Q-learning and SARSA for target search and path planning, while preventing severe power cost under various disturbance conditions is under question. Particularly, in this case we are interested to evaluate the adaptability of the two algorithms to the wind intensity. Figure. \ref{fig: grid_prior}, demonstrates the reward of both algorithms, upon finding the optimal path for each goal across all wind intensities. The reported rewards are an average of accumulated rewards of each episode for the last 10\% of overall episodes prior to the convergence of policy. 

The results suggest that,  goal 1 could be reached with minimal cost via both RL algorithms in a single battery life. However, the head-wind fields $(W_{max}/2,W_{max})$ could cause a fast drop in battery level due to the drag force. When the wind reached to its maximum amount, any path that did not visit the charging spot ended with a termination due to the battery loss. The interesting case however, is for $w_x=W_{max}/2$: SARSA generated the path to visit the charging spot resulting in lower rewards and lower overall cost, in comparison with the path of Q-learning. In contrast, Q-learning found a path to reach the goal 2 without recharging. If we look at other accumulated rewards, often SARSA generated a more conservative path. The reason behind this discrepancy is the greedier nature of Q-learning. Once the goal state is found, the value of state-action pairs that lead to the goal state increases due to the embedded maximization, in value update rule (Eq. \ref{eq: sars}). This updated value for the particular adjacent goal states propogates to other cells as the learning proceeds, resulting in the obtained path.

The task in hand for real-world application requires exploration in large areas when often no positive reward can be found for the duration of an episode. This is known as a common challenge where the RL agent should explore in an environment with sparse goals. In particular, for the task with the maximum head-wind it was observed that, moving the charging sport from where it is in Fig. \ref{fig: baseline} to $(x_c,y_c)=(20,0)$ can increase the required number of episodes by a factor of 10 for reliable convergence. The exploration strategy was shown to be the prominent factor in the case of discovering the charging location. $\epsilon$-greedy algorithm is a common exploration technique for policy improvement \cite{sutton2018reinforcement}. Nevertheless, a high-level supervision on the exploration-exploitation trade off can play a prominent role in the convergence of the RL algorithms. Therefore, the $\epsilon_{episodic}$ (Eq. \ref{eq: episodic}) was proven to improve the convergence rate of the RL agent by enhancing the exploratory actions in initial episodes and relying on the policy for the later episodes. For the scenarios with multiple goals in the same environment (no new instance of wind field, degradation factor or power model) we witnessed a significant improvement in convergence rate by updating $\epsilon_{episodic}$ exponentially once an estimate of overall number of required episodes is available.

Ascend in camera's altitude normally constitutes a larger captured frame. However, it is evident that this ascend usually results in a compromise on the performance of typical on-board cameras and the associated object detection frameworks. Here we are trying to understand, if the path planner algorithm can adjust to the environmental (including the vehicles hardware) shortcomings. Particularly in this simulation environment, the upper altitude range is set high enough, so that color segmentation algorithm fails to detect some objects. This directly will be reflected in the reward function which demotivates the vehicle to rise too high, as the extremely high altitudes lead to fewer detected objects and lower rewards. On the other hand, due to the presence of the adaptive-$\epsilon$ policy, the agent will strive to explore the available $s_z$ states. This exploration comes at the cost of a slower convergence, as the termination in many episodes will occur due to the depletion of the battery in the newly explored states.

%-----------------------------------------------------------------------
\subsection{3-Dimensional path planning}
\begin{figure}[h]
  %\scalebox{.5}{\input{plot.tex}}
  \centering
  \includegraphics[width=10cm]{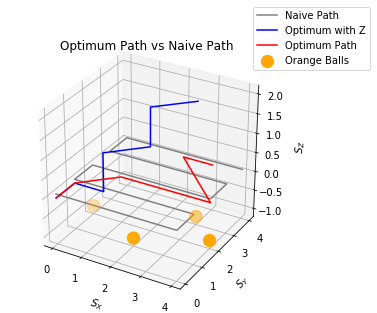}  
  \caption{A demonstration of generated path by the RL planner with and without movement in z axis. The 3D movement requires only 5 steps to detect all the objects compared to the planar motion which requires 7 steps.}
  \label{fig: 3Dmovement}
%  \FloatBarrier
\end{figure}

\begin{figure}[]
  %\scalebox{.5}{\input{plot.tex}}
  \centering
  \includegraphics[width=10cm]{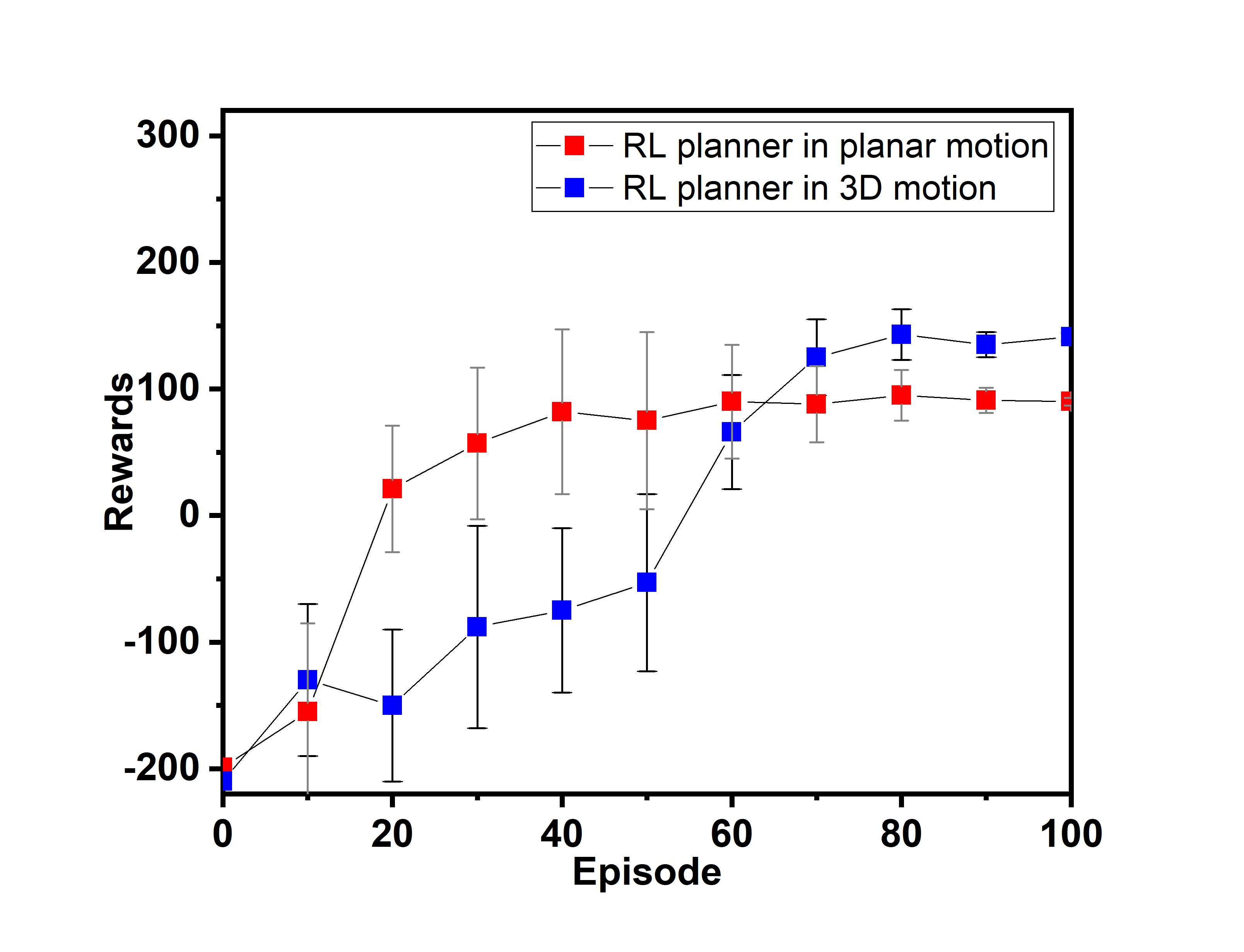}  
  \caption{The collected rewards by RL planner in 3D and planar motion averaged across 5 trials with various object distribution and disturbances.}
  \label{fig: 3DnPlanar}
%  \FloatBarrier
\end{figure}

Fig. \ref{fig: 3Dmovement} demonstrates a comparison between a RL planner which can chose the actions 9 and 10 to move along the z axis. In order to compare the performance of the algorithms, a $630 m \times 630 m$ environments with 10 objects were designed. The objects will be relocated every 100 episodes at random spots and new wind condition will be drawn from the wind state list ($w_x\in[-W_{max},  W_{max}]$). Fig. \ref{fig: 3DnPlanar} demonstrates the average of collected rewards of the RL model in 3D and planar motion across 10 episode intervals across 5 various objective distribution. It can be seen that, operating in 3D motion will result in higher rewards, or more detected objects per battery life but the algorithm convergence significantly slower. It should be noted that, the conclusion drawn by this experiment may change depending on the object detection technique itself. This divergence may expand when the applied to pre-trained CNNs for drones. The idea is to enable robust adaptability of the learning framework depending on the operational shortcomings. The behavioral policy that is developed in the mental rehearsal stage throughout the simulations, is needed to inevitably adapt based on the end-point operational needs. This can include the change in power cost of the quadrotor (varying from day to day), and the failure of the RGB or Lidar sensor depending on rain and other uncertainties as repeatedly appears in the literature\cite{8206048,8335812}.

%-----------------------------------------------------------------------

%--------------------------------------------------------------
%--------------------------------------------------------------
\subsection{Energy-aware exploration}
\begin{figure}[]
  %\scalebox{.5}{\input{plot.tex}}
  \centering
  \includegraphics[width=15cm]{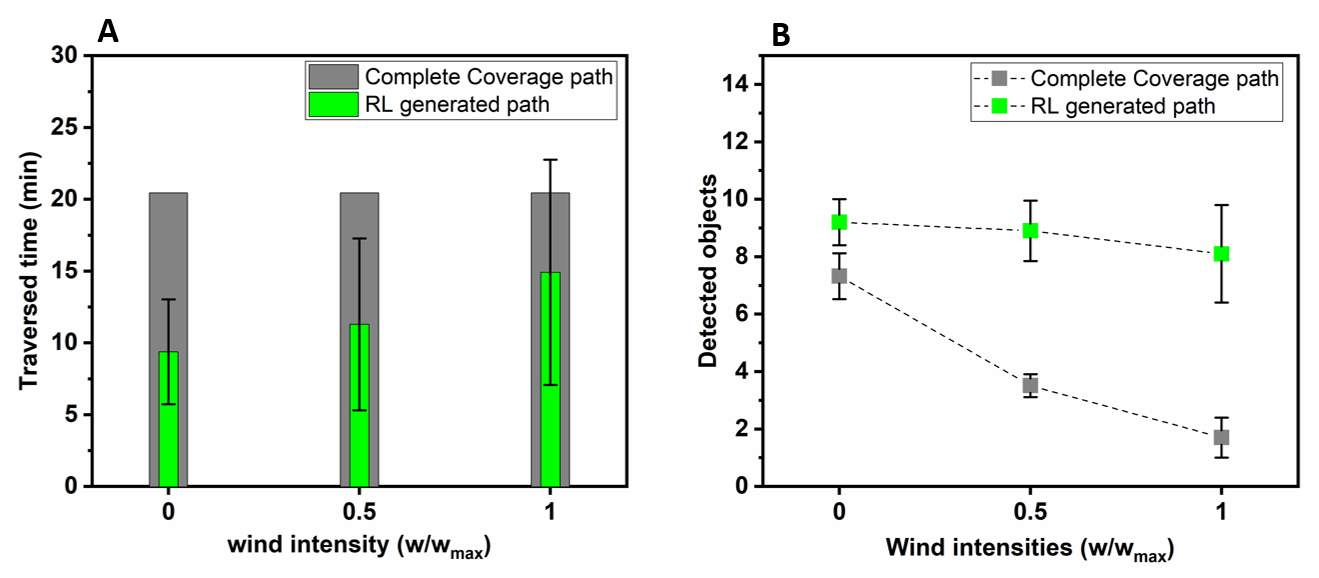}  
  \caption{A comparison between the performance of the RL-generated path and complete coverage path for detecting 10 randomly distributed goal objects under three normalized wind intensities. (A) The total required travelling time for each path planner to detect all the target objects, with unlimited battery assumption. (B) Number of detected objects within each battery life.}
  \label{fig: traversed_time}
%  \FloatBarrier
\end{figure}

Learning from power model, can demonstrate its merits during exploration tasks in large outdoor areas. In such cases, complete coverage of the entire search domain may not be possible within a limited battery life. On the other hand, detecting the desired objects may be achievable in extremely shorter flight duration. Fig.~\ref{fig: traversed_time} demonstrates a comparison between the performance of RL generated path and complete coverage counterpart, for the task of detecting 10 randomly distributed goals.  

We first ignored the battery limitations on the UAV's battery life, and measured the required traversed time after 10 learning episodes. The results demonstrate the resultant flight duration across 5 experiments (5 different goal distribution) and 3 wind intensities that are normalized with respect to $w_{max}$. The search domain size is adjusted so that a quadrotor can sweep the entire domain with $50\%$ overlap in a common SOTA flight time of 20 minutes. As it is expected, when there is no wind, the optimal path is rapidly found by the RL agent. However, the deviation of the flight duration grows under higher wind intensities, since the agent is highly discouraged by the value matrix to resist the wind and thus will stick to greedier actions with lower negative rewards, resulting in a longer flight time.
The more interesting outcome can be found by looking at the performance of the two path planning algorithms within a limited battery life (Fig~\ref{fig: traversed_time} right). From the drag dataset (Fig.~\ref{fig: DragTheta}) we observed that the motion of the UAV with respect to wind can demonstrate a highly nonlinear power cost. Therefore, in cases with high intensities the complete coverage path often results in rapid depletion of the battery. In an average wind field ($w/w_{max} =0.5$), the energy-aware RL planner demonstrates a more than 2-fold improvement in the object detection performance.

The notion of exploration strategy is a key factor for both scenarios 2 and 4. In scenario 4, the state-space alone consists of 2205 which constitutes to state-action space of 110250 when solved for 5 wind fields and 3D actions are allowed. Thus, training the model in the simulation prior to transferring the action values appears to be essential. In summary, both the Goal selection and the path planning problem can benefit from the proposed RL framework in terms of the operational time of the quadrotor on a single battery while accomplishing the object detection task. The adaptive framework can pave the way for UAV operations with exploratory objectives such as vegetation detection, and target search and rescue in vast areas.

\section{Conclusion}
The idea of visual exploration with autonomous vehicles is an attractive topic, yet bearing many challenges. Energy-aware goal selection and path planning of such vehicles via reinforcement learning was addressed here.
The challenge that was tackled is the presence of disturbance at the time of operation, which influences the performance of the robot both in terms of movement cost and the reliability of the computer vision components.
The fundamental idea of this work revolves around incorporation of the reward from the detected goal objects and the cost from the movement of the agent. The proposed algorithm appeared to be particularly applicable, for the missions with spars goals where the agent is constrained by its limited battery life. The ability of a properly tuned RL-based agent to learn the effect of newly emerged wind field suggests the applicability of this algorithm for fully autonomous task completion in large fields. The capability of this framework to adapt to new wind fields and consequent power consumption model through value matrix updates suggests facile transfer of the developed policy from one flying robot to another. 
It was shown that, the classic Q-learning framework is capable of minimizing the power cost of UAV navigation while outperforming the complete coverage solutions by 2-folds in an average wind field. Although an improvement on $\epsilon$-greedy exploration paradigm was incorporated here, there appears to be significant limitations at the time of exploration for the Q-learning algorithm. Additionally, the use of cameras with other heading angles are proved to be helpful and is considered for future work.

\bibliographystyle{unsrt}
\bibliography{references}

\begin{thebibliography}{10}

\bibitem{sun2016camera}
Jingxuan Sun, Boyang Li, Yifan Jiang, and Chih-yung Wen.
\newblock A camera-based target detection and positioning uav system for search
  and rescue (sar) purposes.
\newblock {\em Sensors}, 16(11):1778, 2016.

\bibitem{vasisht2017farmbeats}
Deepak Vasisht, Zerina Kapetanovic, Jongho Won, Xinxin Jin, Ranveer Chandra,
  Sudipta Sinha, Ashish Kapoor, Madhusudhan Sudarshan, and Sean Stratman.
\newblock Farmbeats: An iot platform for data-driven agriculture.
\newblock In {\em 14th $\{$USENIX$\}$ Symposium on Networked Systems Design and
  Implementation ($\{$NSDI$\}$ 17)}, pages 515--529, 2017.

\bibitem{koch2019reinforcement}
William Koch, Renato Mancuso, Richard West, and Azer Bestavros.
\newblock Reinforcement learning for uav attitude control.
\newblock {\em ACM Transactions on Cyber-Physical Systems}, 3(2):1--21, 2019.

\bibitem{bezzo2016online}
Nicola Bezzo, Kartik Mohta, Cameron Nowzari, Insup Lee, Vijay Kumar, and George
  Pappas.
\newblock Online planning for energy-efficient and disturbance-aware uav
  operations.
\newblock In {\em 2016 IEEE/RSJ International Conference on Intelligent Robots
  and Systems (IROS)}, pages 5027--5033. IEEE, 2016.

\bibitem{di2015energy}
Carmelo Di~Franco and Giorgio Buttazzo.
\newblock Energy-aware coverage path planning of uavs.
\newblock In {\em 2015 IEEE international conference on autonomous robot
  systems and competitions}, pages 111--117. IEEE, 2015.

\bibitem{vaddi2019objectDetection}
Subrahmanyam Vaddi, Chandan Kumar, and Ali Jannesari.
\newblock Efficient object detection model for real-time uav applications.
\newblock {\em arXiv preprint arXiv:11906.00786}, pages 1--16, May 2019.

\bibitem{li2006q}
Yibin Li, Caihong Li, and Zijian Zhang.
\newblock Q-learning based method of adaptive path planning for mobile robot.
\newblock In {\em 2006 IEEE international conference on information
  acquisition}, pages 983--987. IEEE, 2006.

\bibitem{panov2018grid}
Aleksandr~I Panov, Konstantin~S Yakovlev, and Roman Suvorov.
\newblock Grid path planning with deep reinforcement learning: Preliminary
  results.
\newblock {\em Procedia computer science}, 123:347--353, 2018.

\bibitem{liu2018energy}
Chi~Harold Liu, Zheyu Chen, Jian Tang, Jie Xu, and Chengzhe Piao.
\newblock Energy-efficient uav control for effective and fair communication
  coverage: A deep reinforcement learning approach.
\newblock {\em IEEE Journal on Selected Areas in Communications},
  36(9):2059--2070, 2018.

\bibitem{wang2020deep}
Hao-nan Wang, Ning Liu, Yi-yun Zhang, Da-wei Feng, Feng Huang, Dong-sheng Li,
  and Yi-ming Zhang.
\newblock Deep reinforcement learning: a survey.
\newblock {\em Frontiers of Information Technology \& Electronic Engineering},
  pages 1--19, 2020.

\bibitem{mammadli_ea:taco:2019}
Rahim Mammadli, Felix Wolf, and Ali Jannesari.
\newblock The art of getting deep neural networks in shape.
\newblock {\em ACM Transactions on Architecture and Code Optimization (TACO)},
  15(4):62:1--62:21, January 2019.

\bibitem{asli2020energyaware}
A.~E.~Niaraki Asli, J.~Roghair, and A.~Jannesari.
\newblock Energy-aware goal selection and path planning of uav systems via
  reinforcement learning, 2020.

\bibitem{levine2016end}
Sergey Levine, Chelsea Finn, Trevor Darrell, and Pieter Abbeel.
\newblock End-to-end training of deep visuomotor policies.
\newblock {\em The Journal of Machine Learning Research}, 17(1):1334--1373,
  2016.

\bibitem{al2012wind}
Wesam~H Al-Sabban, Luis~F Gonzalez, Ryan~N Smith, and Gordon~F Wyeth.
\newblock Wind-energy based path planning for electric unmanned aerial vehicles
  using markov decision processes.
\newblock In {\em Proceedings of the IEEE/RSJ International Conference on
  Intelligent Robots and Systems}. IEEE, 2012.

\bibitem{yue2013ansys}
Chunfeng Yue, Shuxiang Guo, and Maoxun Li.
\newblock Ansys fluent-based modeling and hydrodynamic analysis for a spherical
  underwater robot.
\newblock In {\em 2013 IEEE International Conference on Mechatronics and
  Automation}, pages 1577--1581. IEEE, 2013.

\bibitem{chovancova2014mathematical}
Ane{\v{z}}ka Chovancov{\'a}, Tom{\'a}{\v{s}} Fico, L'ubo{\v{s}} Chovanec, and
  Peter Hubinsk.
\newblock Mathematical modelling and parameter identification of quadrotor (a
  survey).
\newblock {\em Procedia Engineering}, 96:172--181, 2014.

\bibitem{zamora2016extending}
Iker Zamora, Nestor~Gonzalez Lopez, Victor~Mayoral Vilches, and
  Alejandro~Hernandez Cordero.
\newblock Extending the openai gym for robotics: a toolkit for reinforcement
  learning using ros and gazebo.
\newblock {\em arXiv preprint arXiv:1608.05742}, 2016.

\bibitem{chen2012evaluating}
Ian Yen-Hung Chen, Bruce MacDonald, and Burkhard W{\"u}nsche.
\newblock Evaluating the effectiveness of mixed reality simulations for
  developing uav systems.
\newblock In {\em International conference on simulation, modeling, and
  programming for autonomous robots}, pages 388--399. Springer, 2012.

\bibitem{sutton2018reinforcement}
Richard~S Sutton and Andrew~G Barto.
\newblock {\em Reinforcement learning: An introduction}.
\newblock MIT press, 2018.

\bibitem{sanders2016introduction}
Andrew Sanders.
\newblock {\em An introduction to Unreal engine 4}.
\newblock CRC Press, 2016.

\bibitem{shah2018airsim}
Shital Shah, Debadeepta Dey, Chris Lovett, and Ashish Kapoor.
\newblock Airsim: High-fidelity visual and physical simulation for autonomous
  vehicles.
\newblock In {\em Field and service robotics}, pages 621--635. Springer, 2018.

\bibitem{maciel2019online}
Bruna~G Maciel-Pearson, Letizia Marchegiani, Samet Akcay, Amir
  Atapour-Abarghouei, James Garforth, and Toby~P Breckon.
\newblock Online deep reinforcement learning for autonomous uav navigation and
  exploration of outdoor environments.
\newblock {\em arXiv preprint arXiv:1912.05684}, 2019.

\bibitem{torres2015automatic}
Jorge Torres-S{\'a}nchez, Francisca L{\'o}pez-Granados, and Jos{\'e}~M
  Pe{\~n}a.
\newblock An automatic object-based method for optimal thresholding in uav
  images: Application for vegetation detection in herbaceous crops.
\newblock {\em Computers and Electronics in Agriculture}, 114:43--52, 2015.

\bibitem{lyu2017autonomous}
Pin Lyu, Yasir Malang, Hugh~HT Liu, Jizhou Lai, Jianye Liu, Bin Jiang, Mingzhi
  Qu, Stephen Anderson, Daniel~D Lefebvre, and Yuxiang Wang.
\newblock Autonomous cyanobacterial harmful algal blooms monitoring using
  multirotor uas.
\newblock {\em International journal of remote sensing}, 38(8-10):2818--2843,
  2017.

\bibitem{8206048}
S.~{Bohez}, T.~{Verbelen}, E.~{De Coninck}, B.~{Vankeirsbilck}, P.~{Simoens},
  and B.~{Dhoedt}.
\newblock Sensor fusion for robot control through deep reinforcement learning.
\newblock In {\em 2017 IEEE/RSJ International Conference on Intelligent Robots
  and Systems (IROS)}, pages 2365--2370, 2017.

\bibitem{8335812}
H.~C. {Oliveira}, V.~C. {Guizilini}, I.~P. {Nunes}, and J.~R. {Souza}.
\newblock Failure detection in row crops from uav images using morphological
  operators.
\newblock {\em IEEE Geoscience and Remote Sensing Letters}, 15(7):991--995,
  2018.

\end{thebibliography}
\end{document}